\documentclass[aps,pre,reprint,superscriptaddress,floatfix]{revtex4-2}

\usepackage{graphicx}% Include figure files
\usepackage{dcolumn}% Align table columns on decimal point
\usepackage{bm}% bold math
\usepackage{fourier}
\usepackage[T1]{fontenc}

\usepackage[normalem]{ulem}
\usepackage[hidelinks]{hyperref}
\hypersetup{
  colorlinks=true,
  allcolors=blue,
  urlcolor=black
  }
\def\*#1{\boldsymbol{\mathbf{#1}}}

\begin{document}

\preprint{APS/123-QED}

\title{Direct observation of small scale capillary wave turbulence using high speed digital holographic microscopy}% 

\author{William Connacher}
 \affiliation{
Medically Advanced Devices Laboratory, in the Center for Medical Devices\\
Department of Mechanical and Aerospace Engineering, Jacobs School of Engineering\\
Department of Surgery, School of Medicine\\
University of California San Diego, La Jolla, CA 92093, USA
}
\author{Jeremy Orosco}
 \affiliation{
Medically Advanced Devices Laboratory, in the Center for Medical Devices\\
Department of Mechanical and Aerospace Engineering, Jacobs School of Engineering\\
Department of Surgery, School of Medicine\\
University of California San Diego, La Jolla, CA 92093, USA
}
\author{Oliver Schmidt}
 \affiliation{Department of Mechanical and Aerospace Engineering, Jacobs School of Engineering, University of California San Diego, 9500 Gilman Drive, La Jolla, CA 92019 USA}
\author{James Friend}
 \email{jfriend@ucsd.edu}
 \homepage{http://friend.ucsd.edu}
 \affiliation{
Medically Advanced Devices Laboratory, in the Center for Medical Devices\\
Department of Mechanical and Aerospace Engineering, Jacobs School of Engineering\\
Department of Surgery, School of Medicine\\
University of California San Diego, La Jolla, CA 92093, USA
}

\date{\today}

\begin{abstract}
It is now known that capillary waves driven upon a fluid interface by high frequency ($>1$~MHz) ultrasound exhibit capillary wave turbulence: the appearance of waves with phase and wavelength far removed from the excitation signal that drives them. An important step towards understanding atomization phenomena driven in this system, these capillary waves may now be studied using high-speed digital holographic microscopy. We observe Zakharov-Kolmogorov weak wave turbulence for a limited range of input power, and find broader turbulence phenomena outside this range. We see discrete thresholds as the input power is increased, where higher and higher frequency responses are driven in the capillary waves with sudden onset between regimes. Here, we employ spatial analysis to find one such extension of the capillary wave response to higher frequencies, suggesting there is additional information in the spatial distribution of the capillary wave that is rarely if ever measured. We verify via frequency modulation that nonlinear resonance broadening is present, which undermines the use of Faraday wave or parametric wave theories to characterize these waves, important in the context of atomization which is not a Faraday wave process.
\end{abstract}

\keywords{acoustofluidics, capillary waves, capillary wave turbulence, Zakharov-Kolmogorov turbulence, nonlinear resonance broadening, Faraday waves, atomization}

\maketitle

% ---------------------------------------------- %
% Intro
% ---------------------------------------------- %
\section{\label{intro}Introduction}

Weakly nonlinear interactions between a large number of waves with random phase result in wave turbulence that is typically modeled with the well developed statistical theory, `weak' wave turbulence (WWT)~\cite{newell_wave_2011}. Ample experimental work has been done on wave turbulent systems including liquid surface waves~\cite{falcon_capillary_2009,brazhnikov_observation_2002}, plasmas~\cite{robinson_scalings_1996,yoon_generalized_2000}, and solid plates~\cite{boudaoud_observation_2008}. Studies in this area frequently provide evidence in support of WWT theory. A fundamental result of WWT is the Kolmogorov-Zakharov (KZ) spectrum, which shows the power spectral density (PSD) of the waves has a power-law dependence upon the frequency (or wavenumber) over a conservative regime of the wave response. For liquid surface waves dominated by surface tension (\emph{i.e.},, capillary waves), the KZ spectrum is represented by
\begin{equation}
    S(f) \propto \epsilon^{1/2} \Big(\frac{\gamma}{\rho}\Big)^{1/6} f^{-\alpha},
\end{equation}
where $\epsilon$ is the mean energy flux, $\gamma$ is the surface tension, $f$ is the frequency, $\rho$ is the liquid density, and $\alpha = 17/6$ for capillary waves~\cite{zakharov_weak_1971,zakharov_kolmogorov_1992}. In a log-log plot of the PSD, the power law produces a PSD that is linearly dependent upon $-\alpha$. In this model, energy is said to `cascade' from low to high frequency in the system with the predicted scaling when certain conditions are met: i) the domain is infinite, ii) there is sufficient scale separation between energy injection and energy dissipation, iii) the cascade is driven by weakly nonlinear three-wave interactions, and iv) the wave interactions are local (\emph{i.e.},, they occur between waves with similar wavelengths).

Recent studies have attempted to reconcile discrepancies between the idealized conditions of WWT and the results of experiments where these idealizations are routinely violated. \citet{connaughton_discreteness_2001} put forward a simple model that explains how quasi-resonances in finite domain capillary wave systems cause deviations from the predictions of WWT. At low levels of nonlinearity, discrete resonances are not broad enough to permit energy to traverse the eigengrid spacing imposed by the bounded domain, and so the energy cascade typical of WWT is stunted. \citet{falcon_observation_2007} observed that the frequency scalings in a gravity-capillary wave turbulence system are dependent on the input power. \citet{falcon_observation_2011} observed that the depth of the liquid also has an impact on the capillary wave spectrum, causing it to deviate from a power law, but the mechanism remains unknown. \citet{deike_energy_2014} measured the wave height spectrum with varying viscosity and showed in real systems that dissipation, indicated by a deviation in the linear spectral slope, occurs in the (theoretically energy conservative) inertial region, between energy injection and the dissipation zone. They propose a new way to measure energy flux in the system and use it to account for non-ideal dissipation. We have previously investigated the effect of increasing nonlinearity beyond what can be considered weak and also showed experimental evidence of finite domain effects~\cite{Orosco:2023aa}. In the current work, we investigate how energy traverses length scales in a capillary wave system with a data-driven approach, taking into account much smaller scales than have been previously studied.

We are primarily interested in milli to micro-scale capillary waves because of their connection to ultrasonic atomization phenomena. Vibration of a surface in contact with liquid in the kHz or MHz range above a threshold amplitude produces many small droplets, on the order of microns, from the liquid surface~\cite{kurosawa_surface_1995,collignon_improving_2018,lang_ultrasonic_1962}. In the kHz range, the size of the resulting droplets can be related to Faraday wave theory, where the driving frequency produces capillary waves at half the driving frequency. \citet{lang_ultrasonic_1962} showed experimentally that using this Faraday wave assumption along with Kelvin's equation relating frequency to wavelength yields a good estimate of the median droplet diameter, $D=\kappa((8 \pi \gamma)/(\rho F^2))^{1/3}$, where $\kappa=0.34$ is a fitting parameter, $\gamma$ is surface tension, $\rho$ is density, and $F$ is driving frequency. This relationship holds for low atomization rates ($\sim 0.01$~mL/s) and ultrasound frequencies between 10-800~kHz. Outside this frequency range, the value of $\kappa$ has been changed post-hoc to $\kappa=3.8$ \cite{kurosawa_characteristics_1997} in an attempt to fit the equation to the results when the frequency was increased to 20~MHz. Post-hoc reasoning was used even for the original definition of $\kappa=0.34$: a wave crest represents one half of a wavelength and therefore it makes sense that $\kappa<0.5$. Lang photographically verified that the capillary waves appearing on the surface form a uniform lattice with wavelength that decreases with increasing ultrasound frequency. Subsequent studies have demonstrated that the droplet size also depends on viscosity, power input, and flow rate with more detailed empirical correlations~\cite{ramisetty_investigations_2013,qi_interfacial_2008,rajan_correlations_2001}. Lang was not able to explain the measured variation in diameter about the median.

Many of these classical conceptions break down in the context of MHz-order, high power ultrasonic atomization. A fundamental assumption of Faraday wave theory---that the excitation frequency is on the same order as the principal capillary wave frequency---fails at frequencies beyond the 100~kHz range. In these small sessile droplet systems, the first oscillations that appear are on the order of 100~Hz, three orders of magnitude lower than the excitation frequency. Moreover, one would expect to see a capillary wave response at one-half the excitation frequency for it to be a Faraday wave. It has been shown experimentally that no peak in the spectrum exists at or near one-half the MHz-order driving frequency~\cite{blamey_microscale_2013}. In fact, a broadband spectrum wave, indicating WWT, develops on the surface at powers far below the threshold for atomization. In this context, Lang's simple approach of deducing droplets from wavelengths becomes untenable because there is no single frequency at which capillary waves occur---they no longer appear in a uniform lattice. The actual droplet size distributions measured from MHz-order ultrasound have two and sometimes three peaks and often disagree with Lang's equation~\cite{collins_atomization_2012,barreras_transient_2002,kooij_size_2019,winkler_saw-based_2015}.

%%%%%%%%% CAVITATION %%%%%%%%%%%
%Cavitation seems to play a role in addition to capillary waves, especially at low frequency~\cite{mir_cavitationinduced_1980,ramisetty_investigations_2013}. It is also important to mention that cavitation becomes less likely at these frequencies, the threshold pressure for cavitation increases and there is a smaller range of compatible bubble sizes\cite{apfel_gauging_1991}. At the frequency used in this work, 7 MHz, the threshold, 0.7 MPa, is not reached unless the vibration amplitude on the substrate reaches ~15 nm. Also take a look at Barreras 2002 paper. Also consider Taylor instability theory work by Peshkin and Raco 1963.

Recent work claims to have solved the problem of determining the droplet size distribution~\cite{kooij_size_2019}; a Gamma function is fit to droplet size distributions demonstrating that they obey well studied corrugation and ligamentation processes found in sprays. This helps to explain the variation about the median that Lang could not explain. However, the Gamma function takes two parameters, the width of the ligament distribution and the ligament corrugation, so that the method has no predictive power but simply exchanges one unknown, the distribution of capillary waves, for another. The authors suggest that, because they measure two droplet size peaks, then there must be a bi-modal distribution of capillary wavelengths on the surface of the liquid. This makes intuitive sense based on a paradigm of Faraday waves and, without measuring waves directly, it is difficult to validate this suggestion. The difficulty associated with gaining direct knowledge of the waves in this context originates in the time and space scales of the waves, which are much smaller and faster than most modern experimental equipment can reliably observe.

In the current work, we study capillary waves experimentally at micro-length and time scales. Our system is a millimeter length, microliter volume of water that completely insulates capillary wave phenomena from the effects of gravity while also clearly expressing finite domain effects. We drive the system using high frequency ($\sim$7~MHz), high power ($\sim$500~mW) ultrasound that pushes the wave interactions up to and beyond the weak nonlinearity assumption. We measure the surface displacement field at 115~kfps, representing a time period of 8.7~$\mu$s, using a bespoke high-speed digital holographic microscope (DHM) capable of producing full field of view measurements of the fluid interface's displacement at up to 120~kfps \cite{emery2021metrology}. 

We first apply standard, single-point time series analysis to this data to obtain standard amplitude spectra. We then analyze the broader spatial data using techniques that compliment and improve upon the results of single-point time series analysis. \citet{berhanu_space-time-resolved_2013} were the first to use this type of data although they were limited to larger length and time scales. They showed that linear and nonlinear dispersion relations extracted from their data matched WWT theories using spatial and temporal Fourier methods. Another commonly used spatial technique is modal analysis using eigenfunction decomposition. The aim is to reveal coherent structures that may not be best represented by sinusoids or wavelets~\cite{taira_modal_2017}. Proper orthogonal decompositions (POD) produce independent spatial modes optimized by energy with no requirement set on time behavior and have been used extensively in turbulence studies~\cite{berkooz_proper_1993,rovira_proper_2021}. We will show that standard modal analyses do not reveal coherent structures in our system. However, they successfully quantify energy shifts across length scales. It is well known that isotropic turbulent systems have POD modes that reduce to a Fourier basis~\cite{taira_modal_2017,schmidt_guide_2020}. The POD modes of our system are indeed closely related to 2D sinusoidal gratings. This detail allows us to attach a well-defined length scale to each mode. Using this approach, we track how energy flows across length scales as a function of power input.

% ---------------------------------------------- %
% Experiment
% ---------------------------------------------- %
\section{\label{exp}Experimental Setup}

\begin{figure*}
\centering\includegraphics[width=\linewidth]{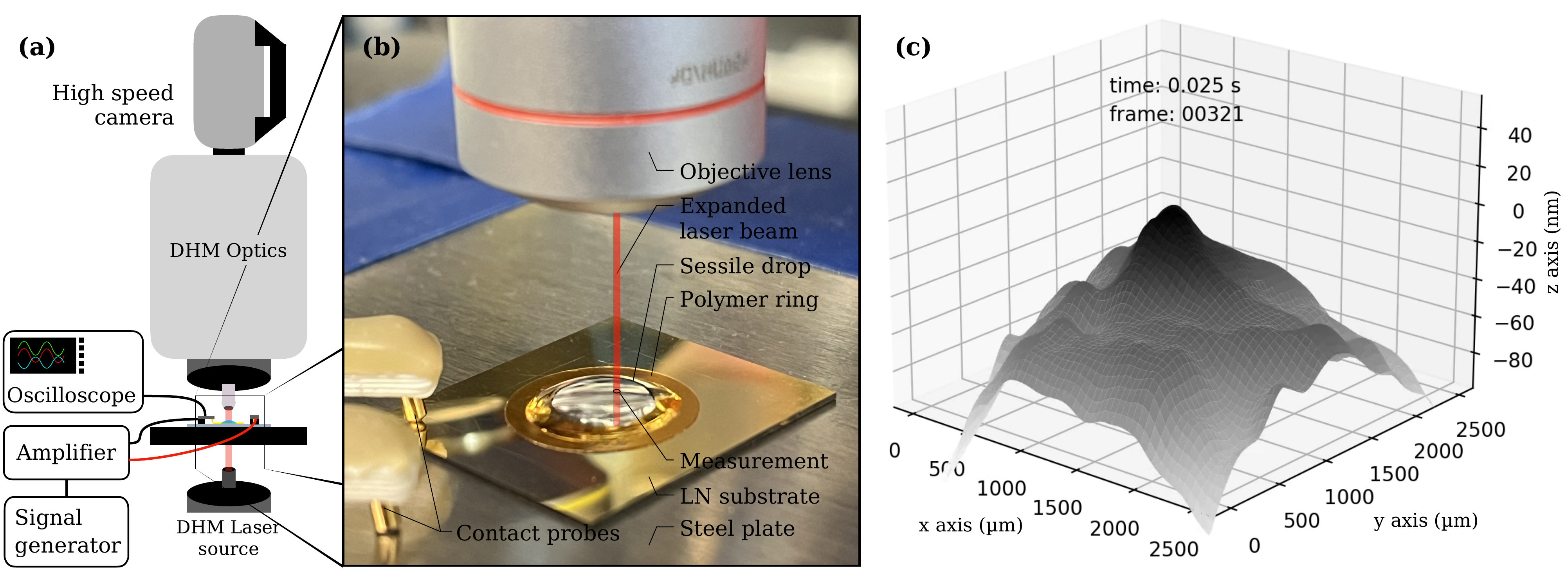}
\caption{\label{fig:experiment} A schematic of our experimental setup incorporating (a) high-speed digital holographic microscopy (DHM) with a (b) sessile droplet placed upon a single-crystal lithium niobate thickness-mode ultrasound resonator. The droplet is placed directly over a transparent ``window'' in the lithium niobate substrate with a gold electrode on both sides everywhere else. An expanded laser propagates from below through the droplet and into the DHM to produce a holographic image captured by the high-speed camera to then produce a (c) height map over the measurement region (indicated in (b) as Measurement). For scale in (b), the lithium niobate resonator is $25\times 25$~mm in size and is 0.5~mm thick.}
\end{figure*}

Our system consists of a sessile drop of water on a thickness-mode $25\times 25$~mm piezoelectric resonator (\emph{see} Fig.~\ref{fig:experiment}). The resonator has a 6-mm diameter circular window in the electrodes that allows light to pass through. The electrodes were formed by sputter depositing a 10-nm titanium bonding layer followed by 500~nm of gold on both sides of a transparent, double-side optically polished ($<0.1$~nm roughness), 0.5-mm thick, 127.68$^\circ$ Y-rotated, X-propagating lithium niobate (LN) wafer. The 36$^\circ$ Y-rotated cut is often used in thickness mode devices, but we have found the 127.68$^\circ$ cut to be superior for operation at resonance \cite{collignon_improving_2018}. A photoresist mask was used on both sides to coat the circular window region with AZ photoresist polymer. After titanium and gold deposition, the  photoresist was dissolved, lifting off the metal films and leaving the transparent LN exposed \cite{Mei:aa}. This procedure was repeated on the second side to produce a thickness-mode resonator with a fundamental resonance at 7~MHz.

A 200~$\mu$m thick with a 9.5~mm inner diameter and a 12.7~mm outer diameter circular polyimide tape (Polyimide Film Electrical Tape 1205, 3M, Maplewood, MN USA) ring was cut and attached to the resonator around the window area in order to repeatably constrain the sessile droplet shape and location during experiments. We used a pipette to place 66~$\mu$L of distilled (ultrapure) water on the LN within the ring so that the drop was pinned around the entire perimeter. To reduce the effects of residual surfactants and contaminants---because these effects cannot be eliminated---the surface was carefully rinsed thrice before conducting each experiment with the ultrapure water. The fluid volume was chosen so that the surface of the drop would be relatively flat compared with the field of view while not deviating significantly from the inherent contact angle of water on LN. All this was sought to ensure the contact line with the ring remains pinned. This geometry also limits the system to deep water capillary waves for the frequencies and wavelengths in this study, so that we can avoid having to consider four-wave interactions in gravity capillary waves as gravity waves and the lateral fluid dynamics from shallow capillary waves \cite{newell_wave_2011}. In fact, very few experiments solely consider capillary waves, as traditional methods for observing such waves require the reduction of gravity through spaceflight, parabolic aircraft flights, or special low-gravity experimental settings \cite{falcon_capillary_2009}. Reducing the apparent gravity, $g$, reduces the critical capillary frequency $f_c\propto g^{(3/4)}$, above which waves are dominated by surface tension. We instead limit the waves that can be generated by reducing the size of the sessile droplet ``container'', thereby preventing most gravity waves and allowing us to focus on high frequency waves.

The resonator was driven directly with a signal generator (WF1946, NF Corporation, Yokohama, Japan), except for the highest power data set which used a linear 10~W amplifier (210L, E\&I, Rochester, NY USA). The signal passed to the resonator was a continuous, fixed amplitude, single frequency sine wave at the fundamental resonance frequency of the thickness mode device. The resonator was placed upon a 1~mm thick, $40\times 40$~mm steel plate with a 10~mm diameter hole aligned with the transparent window in the LN resonator. One spring probe was placed in contact with the top electrode, while another was placed in contact with the steel plate, forming an electrical contact with the bottom electrode. The current and voltage were measured with probes connected to an oscilloscope and their product was averaged over 1 million cycles to determine the true power entering the resonator. The signal was applied to the resonator ~0.5 seconds after beginning measurement with the DHM and maintained until after the measurement was completed at six seconds of elapsed time. Experiments were performed over a range of power inputs with all other conditions controlled. We recorded eighteen data sets between 0--250~mW. The zero power data set was recorded for only two seconds.

We measured the surface displacement in time using a customized digital transmission holographic microscope (DHM, Lynce{\'e}Tec, Lausanne, Switzerland) coupled to an ultra high speed camera (NOVA S12, Photron, San Diego, CA USA). In this approach, laser light, with a wavelength of 660~nm, is split into a sample and a reference beam. The sample beam passes through the window in the resonator and the sessile drop and is then combined with the reference beam to form a phase image and an intensity image at the camera. The phase delay caused by passing through varying distances of water produces a two-dimensional map of the change in the height of the water surface. The measurable change is up to one wavelength of light without post-processing. However, if the height change is sufficiently smooth and gradual, the phase jumps caused by height changes greater than the wavelength of light can be unwrapped to still produce accurate measurements of the fluid interface's deflection. The height resolution of the system is on the order of 10~nm. We refer the reader to \citet{cuche_simultaneous_1999} for details on the DHM technology. %Details on our data processing can be found in the Appendix \TOFIX{set label for appendix}. 
We used a 10X objective lens and recorded at 115,200~frames per second with a $200 \times 200$ pixel field of view, producing a pixel size at the fluid interface of 1.625~$\mu$m. Thus, we expected to observe behavior at frequencies up to 60~kHz and with reasonable amplitude accuracy up to approximately 30~kHz according to the Nyquist–Shannon sampling theorem \cite{Shannon:1949aa}.

% ---------------------------------------------- %
% Time
% ---------------------------------------------- %
\section{\label{time}Time-series analysis}

\begin{figure}
\centering\includegraphics[width=\linewidth]{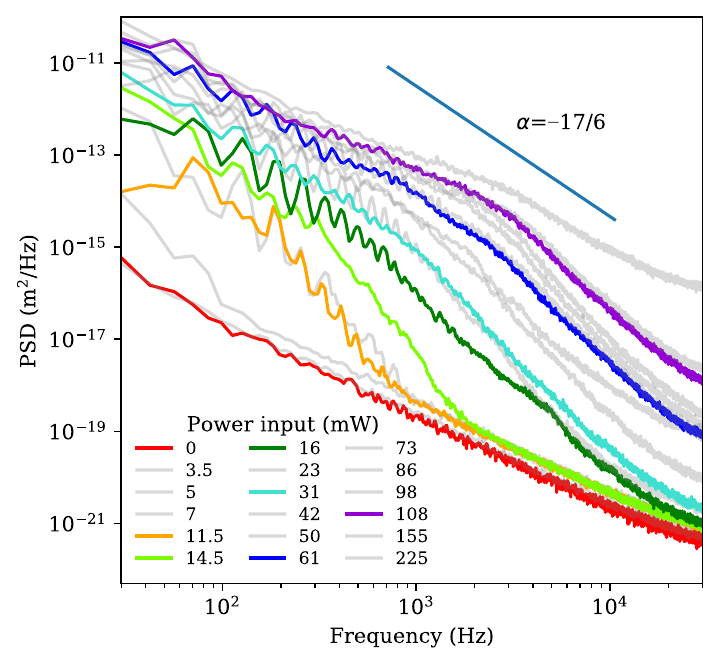}
\caption{\label{fig:FFT} Power spectral density of the interfacial displacement plotted with respect to the frequency for a single point on the fluid interface at 7~MHz excitation. Note the red line indicates no input power, representing the noise floor with our measurement method. As the input power is increased, the appearance of capillary wave oscillations is apparent at about 5--7~mW, and progressive expansion of the cascade to higher frequencies is clear as the power is increased from 14.5~mW to 225~mW. The modal character of the capillary waves is lost from apparent frequency broadening at about 61~mW. The observed spectral slopes are generally lower than the theoretical value of $-17/6$. Colors indicate power input values where there were notable changes in the spectra.}
\end{figure}

In order to obtain the power spectrum, we extracted the data describing the transverse displacement of the  central pixel and performed Welch's method \cite{welch_use_1967} with blocks of $2^{15}$ time steps (\emph{see} Fig.~\ref{fig:FFT} and Appendix A) to minimize noise. Most studies of wave turbulence inject energy at a range of frequencies lower than the expected dissipation range in order to ensure nonlinear wave interactions and the observation of a turbulent cascade~\cite{aubourg_investigation_2016,falcon_observation_2007}. When these systems are driven parametrically, at a single frequency, it is common to observe Faraday waves indicated by a dominant peak at one half the driving frequency and well-ordered patterns in space at low power~\cite{falcon_capillary_2009}. With sufficient power, however, Faraday wave systems do become turbulent~\cite{wright_diffusing_1996}. Our system is fundamentally different, because we drive the system at a frequency five orders of magnitude higher than the frequency of the capillary wave cascade's starting frequency. It is important to note here that the driving signal of 7~MHz cannot be observed in Fig.~\ref{fig:FFT} because the frequency is beyond our high-speed DHM measurement range. However, \citet{zhang_onset_2021} have proposed a mechanism by which large frequency excitation can stimulate much lower frequency wave behavior in a similar system, supported by past data captured by \citet{blamey_microscale_2013} where the spectral response at a single point \emph{was} measured to 25~MHz. The mechanism occurs by the generation of an acoustic standing wave in the parent droplet as a cavity. This produces a spatially varying acoustic pressure upon the fluid interface that causes its deformation. If at low amplitude, the deformation is static \cite{Manor:2011fk}, however, it quickly becomes dynamic beyond an excitation threshold dependent on the specific experimental setup. As the amplitude of the interface's deformation grows larger, the standing acoustic wave in the fluid droplet changes shape, leading to a change in the acoustic pressure upon the interface, changing its shape, and so on to produce capillary wave motion at spatiotemporal scales associated with the 100~Hz range capillary wave resonances in this system. Those capillary wave resonances interact via a nonlinear three-wave mechanism to produce the cascade \cite{newell_wave_2011}.

As has been seen in numerical simulations by Pushkarev and Zakharov, the finite domain in our system fails to produce a cascade at low power inputs where the nonlinearity is very weak~\cite{pushkarev_turbulence_1996}. Based on the model from \citet{connaughton_discreteness_2001}, there are multiple thresholds of resonance broadening and therefore of nonlinearity, all driven by power input. At each of these thresholds, the cascade increases in length towards larger wavenumbers, consistent with experimental observations of the cascade growing to higher frequencies with increases in input power \cite{blamey_microscale_2013}. It seems clear from their work that the specific thresholds for a given system depends on the spacing of the quasi-resonances and the relationship between forcing and broadening.

The amplitude spectra in Fig.~\ref{fig:FFT} clearly show a consistent peak at around 30~Hz with harmonics except for the 0 and 5~mW results. The 0~mW data indicates the noise floor of our measurement system, a combination of shot noise in the measurement system and thermal excitation of the fluid interface. At 3.5, 5, and 7~mW, our system is highly intermittent% as can be clearly seen in a short time Fourier transform of each experiment (\emph{see} Supplementary)
, which explains why the 5~mW line is not separated from the 0~mW line. Up to 11.5~mW the capillary wave energy remains confined to frequencies below about 100~Hz. As the power is increased to 14.5~mW there is a broadening of the harmonics and the beginning of a broadband response that extends the cascade to 200~Hz. At 16~mW the cascade abruptly extends to about 10~kHz, and continues to extend upward in frequency beyond our DHM's measurement range from 31~mW and up. These characteristics seem to support Connaghton's model of finite domain, ``frozen'' turbulence \cite{connaughton_discreteness_2001}.

\begin{figure}
\centering\includegraphics[width=\linewidth]{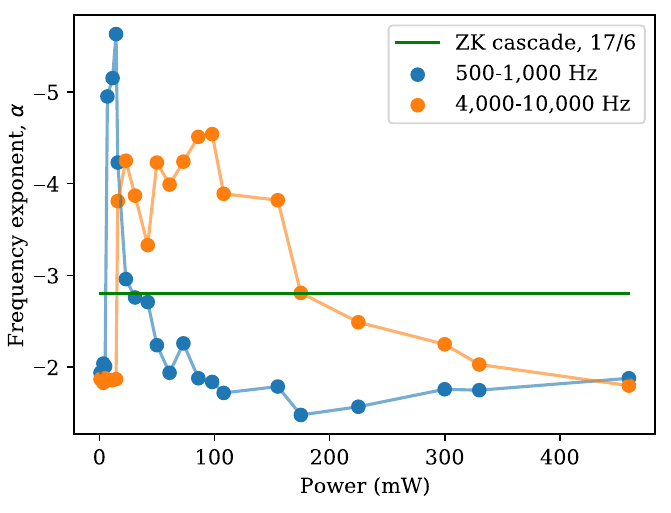}
\caption{\label{fig:slopes} The frequency exponent of the wave height spectra, plotted here as average values determined from the slope of the spectra over the ranges 500--1000~Hz and 4--10~kHz, rarely coincides with the ZK prediction of $\alpha = -17/6$. The low frequency component grows rapidly to more than $-5$ at low powers, a consequence of the appearance of the capillary wave, but shallows to $-2$ as the power increases beyond 50~mW. The change in the high frequency component from $-2$ to $-4$ occurs at the same time the low frequency component shallows from $-5$ to $-2$. As the input power increases from 100~mW, the high frequency component decreases through the ZK prediction to approach a value of $-2$.}
\end{figure}

The predicted value of $\alpha = -17/6$ associated with the Zakharov-Kolmogorov (ZK) capillary wave cascade occurs only in a specific range of power inputs in a specific frequency range, 0.1--1 kHz at 15--35~mW (\emph{see} {Fig.~\ref{fig:slopes}}). The initial jump to large $\alpha$ above the slope of the noise floor at 0~mW in Fig.~\ref{fig:FFT} is a direct result of a finite domain. After the initial cascade extension is complete at around 20~mW, we observe the appearance of a corner frequency separating two regions of constant $\alpha$ at each power input. Below this frequency, the slope $\alpha$ is shallower than the predicted value, and above this frequency it is steeper. This spectral response is the opposite of what is typically observed in systems that contain both gravity and capillary forces driving the waves \cite{falcon_observation_2007}. This would seem to indicate a transition to a stronger dissipation mechanism at the frequency where the shoulder occurs, 1~kHz at 23~mW and monotonically increasing to 4~kHz at 246~mW. Conversely, the low frequency range shifts from $\alpha=-17/6$ just after cascade completion in keeping with WWT towards smaller values of $\alpha$ as power increases. This may indicate a transition from weakly to strongly nonlinear wave interactions consistent with observations of this transition in related experiments \cite{Orosco:2023aa}.

We also observe frequency broadening of the low frequency modes as the power is increased from 16 to 108~mW, in concert with the growth of the high frequency portion of the cascade. The broadening of peaks with increased input power again supports Connaghton's finite domain nonlinear resonance broadening model. Increased nonlinear interactions among discrete resonances allow energy to move down the cascade more easily and may explain faster growth at high frequency.
Notably, we do not see any distinct resonance peaks above 100~Hz at the highest power. This power is well below the threshold of atomization onset for this system, so the presence of capillary waves at a certain frequency seems not to foreshadow the peaks of a droplet size distribution in this system as was suggested by \citet{kooij_size_2019}. 

\begin{figure}
\centering\includegraphics[width=\linewidth]{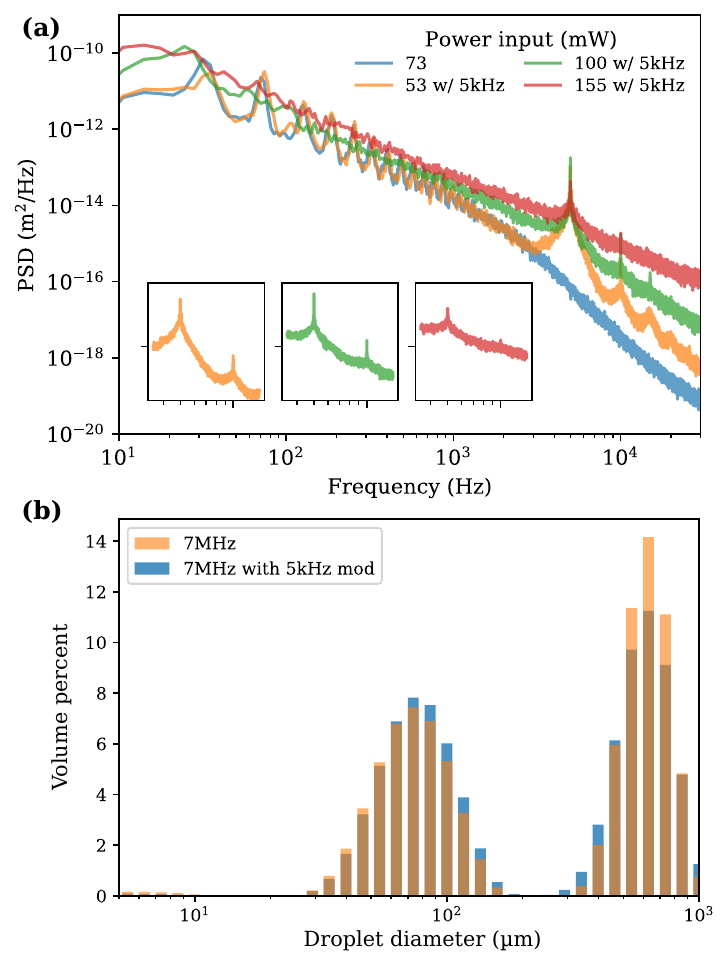}
\caption{\label{fig:modulation} The effect of using 5~kHz modulation of the 7~MHz input signal upon the (a) capillary wave spectrum. At low amplitudes, the effect of the modulation is significant: compare 73~mW without modulation and 53~mW with modulation, and note how the response is significantly perturbed at 5~kHz. At greater amplitudes, the nonlinear resonance broadening appears to \emph{reduce} the amplitude of the 5~kHz peak: compare 155~mW to the 100~mW data, both with the 5~kHz modulation. These input powers are well below what is required for atomization, and there is (b) no significant difference between the droplet sizes that are generated from atomization at 500~mW through the use of signal modulation at 5~kHz.}
\end{figure}

To further explore this issue, we then sought to explicitly modulate the input signal at 5~kHz in an attempt to drive the appearance of a dominant capillary wave at this modulation frequency. At the relatively low power of 53~mW with the sinusoidal modulation of the input frequency, the modulation frequency does appear in the capillary wave with a corresponding strong response peak in Fig.~\ref{fig:modulation}(a), along with three harmonics up to about 20~kHz. As we increase the power to 100~mW, still well below the atomization threshold, the 5~kHz amplitude peak is still strong but broader while the harmonics are much less prominent. Increasing the power to 155~mW further broadens the 5~kHz peak with reduction in its maximum amplitude of about 50\%; the harmonics are completely absent. Increasing the power beyond the measurement capabilities of the DHM to drive atomization at 500~mW, we found that there is no observable difference between the droplet size distribution for the atomized droplets whether or not the 5~kHz modulation is present, as determined using laser diffraction droplet sizing (Spraytec, Malvern Panalytical, Malvern UK) and plotted in Fig.~\ref{fig:modulation}(b).

% ---------------------------------------------- %
% Space
% ---------------------------------------------- %
\section{\label{space}Spatial mode analysis}

Looking beyond the information that can be gleaned from observation of a single point on the fluid interface over time, we next consider the entire field of view and seek to determine how the fluid interface evolves over time \emph{and} space as the input power is changed. We use the machinery of proper orthogonal decomposition to support this effort~\cite{berkooz_proper_1993,rovira_proper_2021}.

\begin{figure*}
\centering\includegraphics[width=\linewidth]{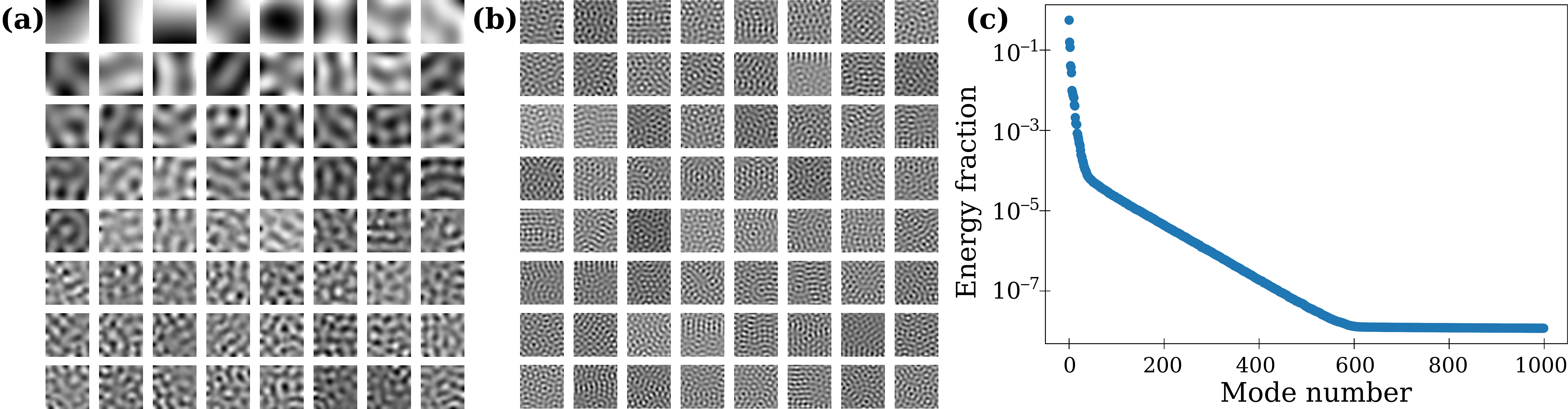}
\caption{\label{fig:PCA} An example of the results from principal component analysis, showing (a) the first 64 modes and (b) the 201$^{\text{st}}$ to 264$^{\text{th}}$ modes for the interfacial displacement from the 16~mW data set. The distribution of the singular value amplitudes in terms of the fraction of the total energy each mode possesses is (c) plotted with respect to the mode number; in this case the vast majority of the response is in the first $\sim 40$ modes and modes beyond about 475 contribute negligibly to the response.}
\end{figure*}

When computing the POD modes and singular values, we selected $2^{13}$ frames from each data set. We transformed each frame into a 40,000 element column vector and collected all the frames together into a $40,000 \times 8,192$ matrix, $F$. The mean frame was subtracted from each frame to produce a matrix, $X$. We then performed a singular value decomposition (SVD) upon $X = U S V^T$, which produced a $40,000 \times 8,192$ matrix $U$, a $8,192 \times 8,192$ matrix $S$, and a third matrix $V$; some details are provided in Appendix B. The matrix $U$ contains 8,192 modes, the diagonal matrix $S$ contains the singular values corresponding to those modes, and $V$ represents the corresponding eigenvectors. By virtue of SVD the singular values are automatically sorted in descending order and are equal to the square root of the eigenvalues, $\lambda$, of the classical eigenvalue problem: $R \bf\phi_i = \lambda_i \bf\phi_i$, where $R$ is the auto-covariance matrix and $\bf\phi$ are the modes \cite{taira_modal_2017}. When the measured data is velocity, the eigenvalues are proportional to the kinetic energy. In our case where the measurement field is displacement, the singular values are therefore proportional to the amplitude scaled by the number of time steps. This means that the most important spatial modes in terms of describing the amplitude of the water surface are the first columns of $U$: they represent the largest displacement components. The modes greater than 475 in each data set were discarded based on the appearance of a prominent corner in the singular value distribution at or below this mode number (\emph{see} {Fig.~\ref{fig:PCA}}), suggesting the modes above 475 play little to no role in defining the displacement of the fluid interface.

For example, the first 64 POD modes and modes 201--264 are plotted in Fig.~\ref{fig:PCA}. The characteristic length scale clearly decreases with increasing mode number. For the wavelength range where the field of view offers sufficient resolution ($<200\,\mu$m), it is evident from 2D FFT (\emph{see} Supplementary Information) that only one length scale is present in each mode, and even for lower modes it is reasonable to assume by inspection that each mode is roughly composed of sinusoids of a single wavelength. In order to quantify the wavelength of the modes, we utilized a custom algorithm based on two-dimensional fast Fourier transforms (2DFFT). For each mode we reshaped its column vector into a $200 \times 200$ image and performed 2DFFT, which produced another $200 \times 200$ image. Each point in this new image is associated with a wavelength and a pixel value associated with the strength of that wavelength. For small wavelengths it is sufficient to define the wavelength as the pixel distance to the pixel with the largest value.  However, for larger wavelengths, poor resolution in the wavelength space means we must take an average of some number of the pixels. Based on trial and error, we chose to average eight pixels with the largest values. Furthermore, in order to eliminate potential bias towards modes that are aligned with the square camera window we performed this procedure on 45 different rotations of the mode image and took the average of the wavelength values over these rotations.

\begin{figure}[!htb]
\centering\includegraphics[width=\linewidth]{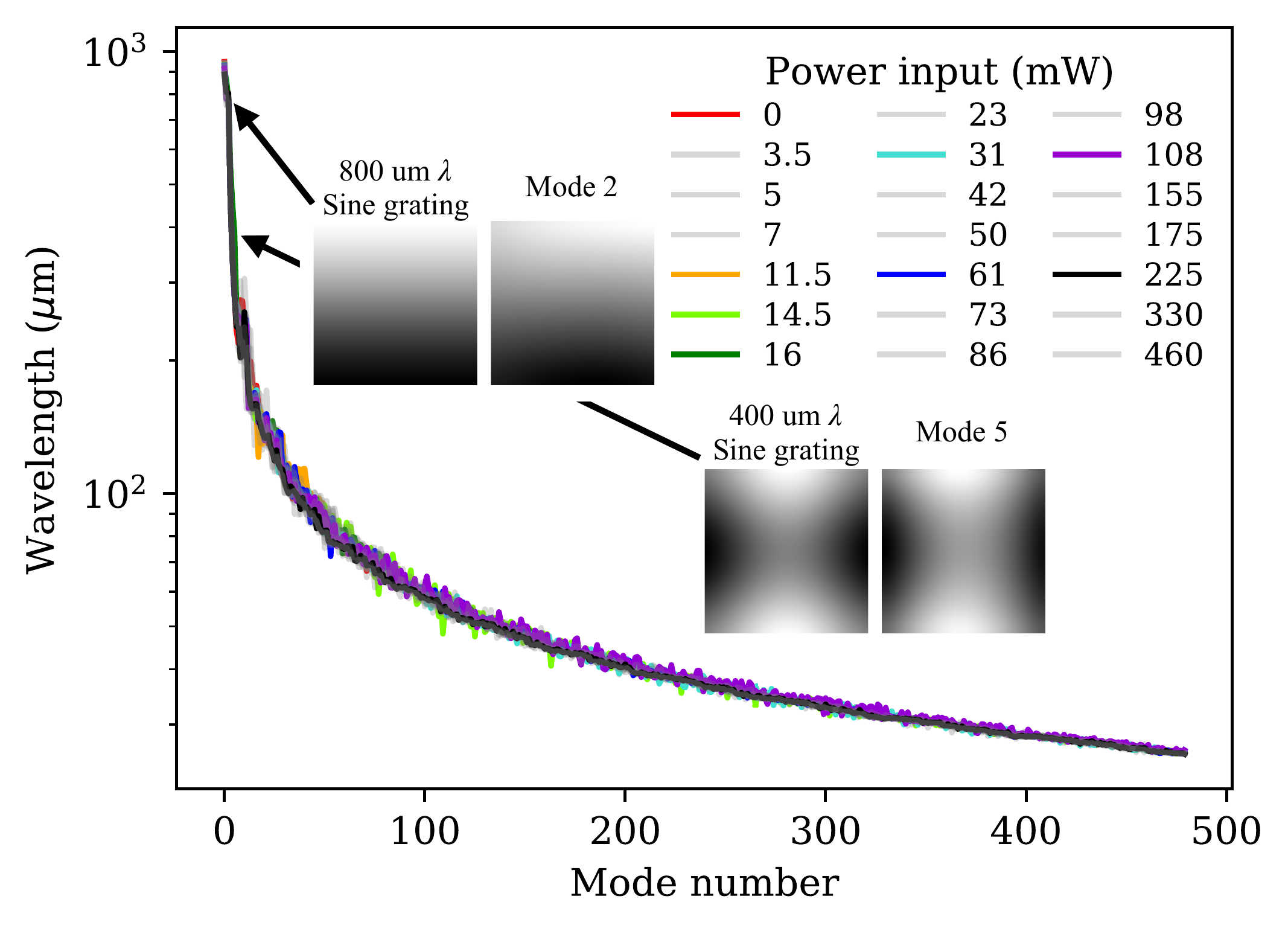}
\caption{\label{fig:WLvsMode} The computed wavelength for each mode found via 2DFFT of the principal component analysis results. The largest measurable modes have a wavelength twice the field of view. Note how the distribution of wavelengths does not significantly change as the input power is increased from 0 to 460~mW.}
\end{figure}

\begin{figure}[!htb]
\centering\includegraphics[width=\linewidth]{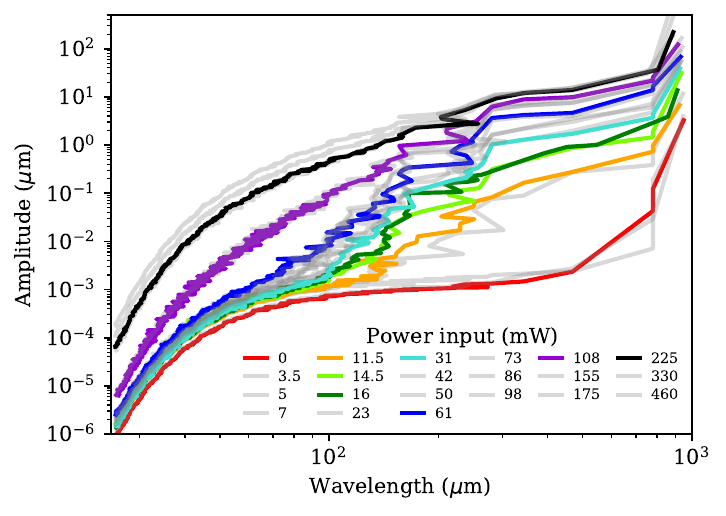}
\caption{\label{fig:SVD} Singular values scaled by the number of frames from POD plotted against the wavelength obtained by performing a 2DFFT algorithm on each mode. This is the same plot as Fig.~\ref{fig:PCA}c), but with the y-axis scaled to be amplitude and with the x-axis transformed from mode to wavelength.}
\end{figure}

In this way, we were able to obtain the relationship between wavelength and mode number in Fig.~\ref{fig:WLvsMode} for the capillary waves in this system at each input power. This allows us to plot the singular values for each mode, which are equal to wave amplitudes as we described earlier, against their respective length scale in Fig.~\ref{fig:SVD}. Energy is clearly restricted to length scales greater than 500~$\mu$m until the power reaches 7~mW. There is also a clear transition between 98 and 108~mW where energy shifts towards much smaller length scales. This second transition was not apparent from Fourier analysis alone (\emph{see} {Fig.~\ref{fig:FFT}}). Increasing power in this range abruptly extends the cascade, as we could not see in Fig.~\ref{fig:FFT} due to frequency resolution, similar to the cascade extension between 7 and 16~mW that \emph{is} observable in Fig.~\ref{fig:FFT}. Finally, notice that while the wave amplitude at scales greater than approximately 200~$\mu$m grow monotonically with an increase in input power, the wave amplitude at smaller length scales grows much more rapidly as the input power increases from 61~mW to 225~mW.

\begin{figure}[!htb]
\centering\includegraphics[width=\linewidth]{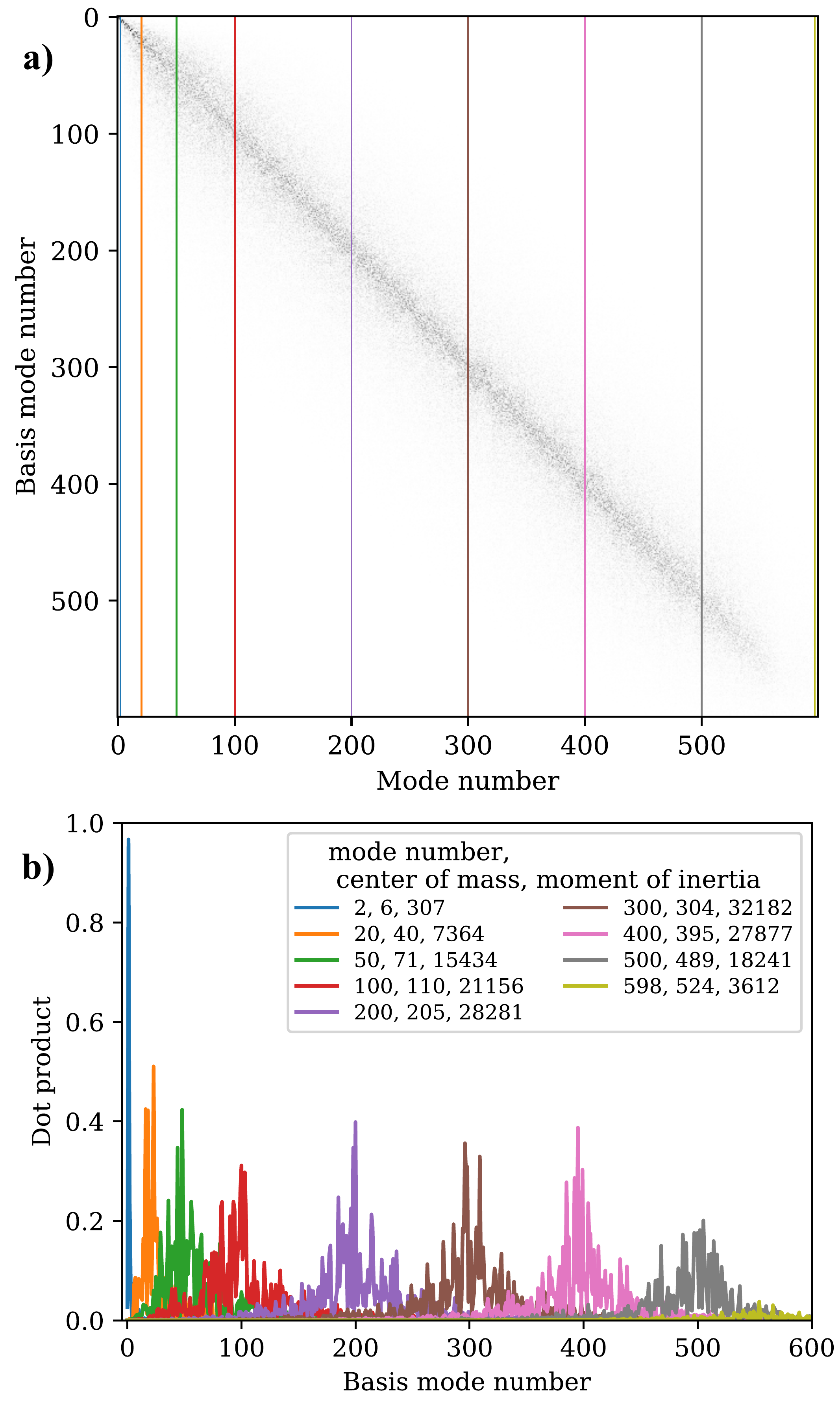}
\caption{\label{fig:Cov} To determine the broadening of the capillary wave modes, we plot the (a) covariance of the 61-mW input power case with respect to the original mode numbers. If there were no broadening of the responses, this would be a diagonal line from upper left to lower right. Instead, it spreads laterally but remains locally distributed as nearly diagonal, indicating that broadening does indeed occur but there are no apparent sub- or super-harmonic parametric resonant cascades across significantly different wavelength scales. The (a) colored vertical lines in the plot are next (b) plotted showing the amplitude of the product of the chosen mode with the basis mode.}
\end{figure}

In order to understand if modes of the same number between data sets are related, we took the covariance between the $U$ matrix for each data set and that of a compiled data set, $U_c$. The compiled data set, comprised of 1024 frames, was constructed from frames taken randomly from each individual data set, upon which the same POD procedure was used to obtain $U_c$. The covariance reveals how similar each $U$ matrix is to a common basis, $U_c$. Figure~\ref{fig:Cov} shows the covariance of $U_{61}$ for the 61~mW case with $U_c$. It also highlights cross-sections (vertical lines) that appear in Fig.~\ref{fig:Cov} where they reveal components of the given mode in terms of the basis modes (\emph{i.e.},, the dot product). For a given mode in an individual data set, the distribution of modal components tend to be centered near the corresponding mode in the basis. We can quantify the deviation of this distribution from the corresponding basis mode using the first and second moments, analogous to the expected value and variance, respectively. The first moment ($\mu_1$) is given by
\begin{equation}
    \mu_1 = \frac{1}{\Delta} \sum_{j=1}^n j D(j),
\end{equation}
where $D(j) = U_{ci} \cdot U_{ij}$ is the distribution of the $i^{\text{th}}$ components of the $j^{\text{th}}$ mode of $U_{ij}$ across the $i$ modes of $U_c$ (e.g.,  the cross-section); and $\Delta = \sum D(j)$ is the sum of these components over $j$. The deviation in the expected value is then the difference between $\mu_1$ and $a$. The second moment, variance $\mu_2$, is given by
\begin{equation}
    \mu_2 = \sum_{j=1}^n D(j) (\mu_1-j)^2.
\end{equation}

\begin{figure}
\centering\includegraphics[width=\linewidth]{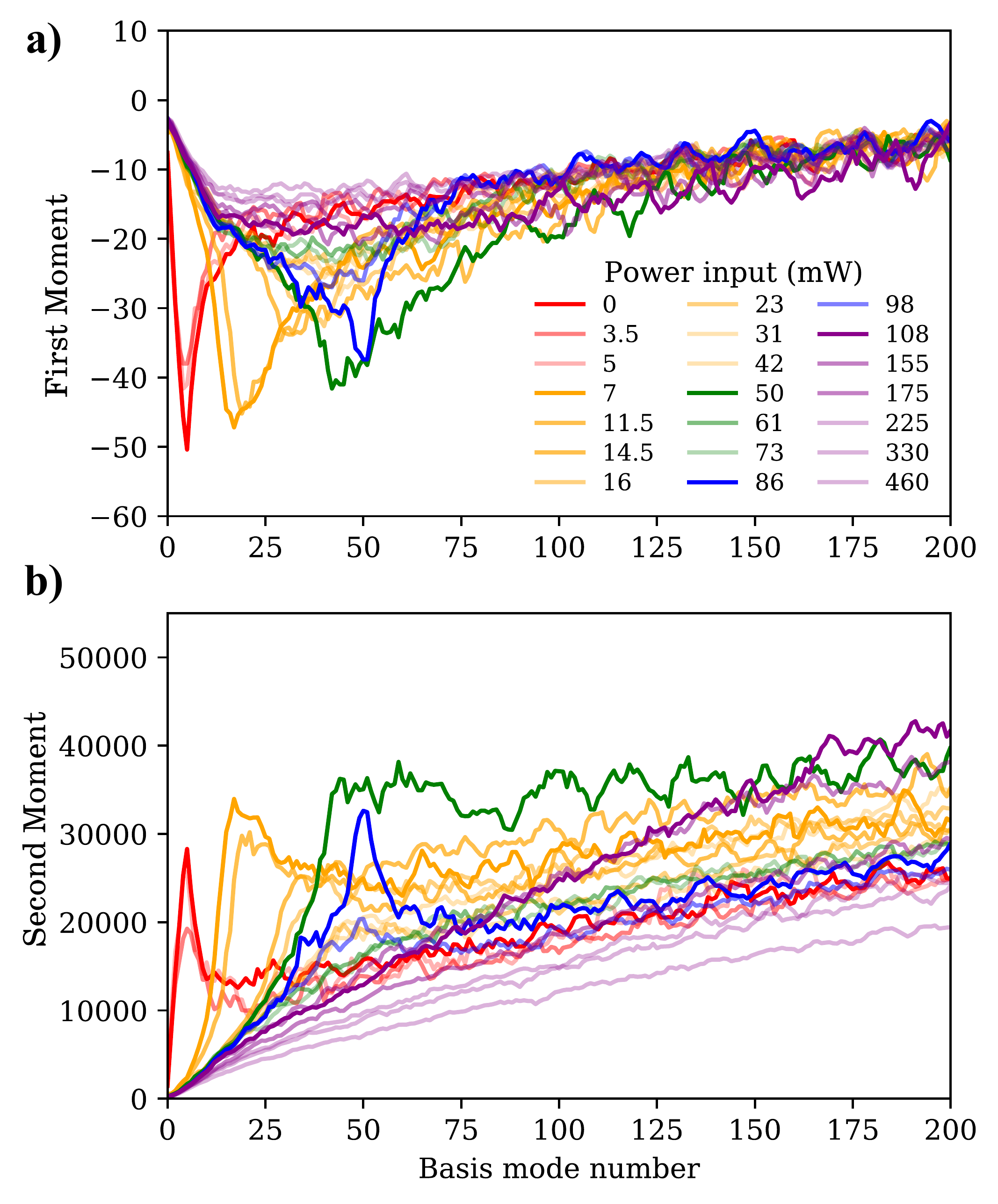}
\caption{\label{fig:CoM-MoI} (a) First and (b) second moments versus basis mode number as a proxy for the wavelength. Recall that the modes are sequentially ordered from largest wavelength to smallest. }
\end{figure}

The variance is a convenient way to show that the modes identified by POD are essentially common between different data sets, not merely just their length scale. They also quantify how the modes themselves change with power input---setting aside the singular values, which rank the amplitudes of the modes within a data set at a single power. We plot the deviation of the first and second moments for each mode of every data set in Fig.~\ref{fig:CoM-MoI}. The first three lines (in red) have a similar shape, but then suddenly the 7~mW line is different, which aligns with the sudden changes in both Fig.~\ref{fig:FFT} and Fig.~\ref{fig:SVD}. The way this shape changes indicates that the modes around the 20th mode are interacting with a much larger number of adjacent modes than they were at only a slightly lower power.

The fact that the cross-sections in Fig.~\ref{fig:Cov} are distributed generally around their corresponding basis mode is an expression of local wave interaction. When the power crosses some threshold near 6~mW, the modes near 20 suddenly become less localized and interact with modes possessing wavelength scales of greater difference. As power increases from 7 to 42~mW the moment lines gradually move towards the zero power line, indicating that the modes interact more locally and they are more closely described by the corresponding basis modes.

Once again, however, at 50~mW the modes beyond mode 45 become de-localized, but again move towards the zero power lines at slightly larger powers. This process repeats from 86~mW for modes near the 50th mode. The final de-localization apparent in the data is centered closer to the 140$^{\text{th}}$ mode, indicating even smaller length scales. This de-localization coincides with the transition identified in Fig.~\ref{fig:SVD}.

% ---------------------------------------------- %
% Conclusions
% ---------------------------------------------- %
\section{\label{conclusions}Conclusions}
Fourier analysis often used in wave turbulence problems identifies the relationship between wave height and frequency for micro-scale capillary waves in a small sessile droplet. The effects of the finite sized droplet produce frozen turbulence and low frequency (relative to the high-speed DHM measurement capabilities) repeated and abrupt cascade extension at specific input powers. Immediately upon cascade completion we see a small region of capillary wave dynamics that correspond to the classic ZK theory of WWT where $\alpha=-17/6$. At greater powers, there is a corner frequency separating $\alpha$ into two frequency domains, a relatively shallow-sloped region at lower frequencies and a steeper-sloped region at higher frequencies. These observations suggest an increasingly strong nonlinear wave interaction at low frequencies and increasing dissipation at larger frequencies.

Spatial analysis with POD provides us with a direct link between amplitude and wavelength. Energy flowing towards smaller length scales exhibits a nonlinear dependence upon the input power, with discrete thresholds repeatedly appearing as the power is increased. These shifts are analogous to cascade extension, most of which are observable using standard Fourier analysis. However, spatial analysis allows us to identify a small scale, high frequency cascade extension that was not apparent with Fourier analysis. The covariance between each data set and the basis modes provides us with information about how coherent modes change with power input. We are able to identify de-localization events where waves interact with wavelengths of greater difference as the power is increased.

Uniform Faraday waves are a poor model for capillary wave dynamics in this system, and therefore are unlikely to be appropriate for predicting atomization which requires greater power. Resonant peaks in the capillary wave response are quickly broadened by the nonlinear coupling, even when the input signal is modulated to force the resonance into existence. This occurs at input power levels well below the atomization threshold, and it remains unclear how to predict which wavelengths will be preferentially amplified---if any---in order to form droplets at a larger power input. A theory of wave turbulence beyond the weak and infinite regimes dictated by WWT are therefore needed to predict which waves will produce droplets in high frequency ultrasound atomization systems.

%In fact, the wavelengths on the surface do not occur in two distinct peaks and there is no evidence of a Faraday wave mechanism. Peaks present at low powers due to geometric resonance or amplitude modulation of the input signal broaden out and disappear at higher powers below the threshold for atomization. This suggests that there must be some yet unknown mechanism that selects certain wavelengths for droplet ejection.

\section{Acknowledgements}
We are grateful to the Office of Naval Research, United States (grants 13423461 and 12368098) and the W.M.\ Keck Foundation, United States for funding provided to J.\ Friend in support of this work. J.\ Orosco is grateful for support provided by the University of California’s Presidential Postdoctoral Fellowship program. We are also grateful to Yves Emery and team at Lynce{\'e}Tec for assistance with adapting the DHM to this project’s needs. Fabrication was performed in part at the San Diego Nanotechnology Infrastructure (SDNI) of UCSD, a member of the National Nanotechnology Coordinated Infrastructure, which is supported by the National Science Foundation (Grant ECCS–1542148).

%\section{Conflict of Interest}
%The authors declare that the research was conducted in the absence of any commercial or financial relationships that could be construed as a potential conflict of interest.

%\section{Publisher’s note}
%All claims expressed in this article are solely those of the authors and do not necessarily represent those of their affiliated organizations, or those of the publisher, the editors and the reviewers. Any product that may be evaluated in this article, or claim that may be made by its manufacturer, is not guaranteed or endorsed by the publisher.

\bibliographystyle{apsrev4-2}
\bibliography{bib}% Produces the bibliography via BibTeX.

\begin{thebibliography}{40}%
\makeatletter
\providecommand \@ifxundefined [1]{%
 \@ifx{#1\undefined}
}%
\providecommand \@ifnum [1]{%
 \ifnum #1\expandafter \@firstoftwo
 \else \expandafter \@secondoftwo
 \fi
}%
\providecommand \@ifx [1]{%
 \ifx #1\expandafter \@firstoftwo
 \else \expandafter \@secondoftwo
 \fi
}%
\providecommand \natexlab [1]{#1}%
\providecommand \enquote  [1]{``#1''}%
\providecommand \bibnamefont  [1]{#1}%
\providecommand \bibfnamefont [1]{#1}%
\providecommand \citenamefont [1]{#1}%
\providecommand \href@noop [0]{\@secondoftwo}%
\providecommand \href [0]{\begingroup \@sanitize@url \@href}%
\providecommand \@href[1]{\@@startlink{#1}\@@href}%
\providecommand \@@href[1]{\endgroup#1\@@endlink}%
\providecommand \@sanitize@url [0]{\catcode `\\12\catcode `\$12\catcode
  `\&12\catcode `\#12\catcode `\^12\catcode `\_12\catcode `\%12\relax}%
\providecommand \@@startlink[1]{}%
\providecommand \@@endlink[0]{}%
\providecommand \url  [0]{\begingroup\@sanitize@url \@url }%
\providecommand \@url [1]{\endgroup\@href {#1}{\urlprefix }}%
\providecommand \urlprefix  [0]{URL }%
\providecommand \Eprint [0]{\href }%
\providecommand \doibase [0]{https://doi.org/}%
\providecommand \selectlanguage [0]{\@gobble}%
\providecommand \bibinfo  [0]{\@secondoftwo}%
\providecommand \bibfield  [0]{\@secondoftwo}%
\providecommand \translation [1]{[#1]}%
\providecommand \BibitemOpen [0]{}%
\providecommand \bibitemStop [0]{}%
\providecommand \bibitemNoStop [0]{.\EOS\space}%
\providecommand \EOS [0]{\spacefactor3000\relax}%
\providecommand \BibitemShut  [1]{\csname bibitem#1\endcsname}%
\let\auto@bib@innerbib\@empty
%</preamble>
\bibitem [{\citenamefont {Newell}\ and\ \citenamefont
  {Rumpf}(2011)}]{newell_wave_2011}%
  \BibitemOpen
  \bibfield  {author} {\bibinfo {author} {\bibfnamefont {A.~C.}\ \bibnamefont
  {Newell}}\ and\ \bibinfo {author} {\bibfnamefont {B.}~\bibnamefont {Rumpf}},\
  }\href {https://doi.org/10.1146/annurev-fluid-122109-160807} {\bibfield
  {journal} {\bibinfo  {journal} {Annual Review of Fluid Mechanics}\ }\textbf
  {\bibinfo {volume} {43}},\ \bibinfo {pages} {59} (\bibinfo {year}
  {2011})}\BibitemShut {NoStop}%
\bibitem [{\citenamefont {Falc{\'o}n}\ \emph {et~al.}(2009)\citenamefont
  {Falc{\'o}n}, \citenamefont {Falcon}, \citenamefont {Bortolozzo},\ and\
  \citenamefont {Fauve}}]{falcon_capillary_2009}%
  \BibitemOpen
  \bibfield  {author} {\bibinfo {author} {\bibfnamefont {C.}~\bibnamefont
  {Falc{\'o}n}}, \bibinfo {author} {\bibfnamefont {E.}~\bibnamefont {Falcon}},
  \bibinfo {author} {\bibfnamefont {U.}~\bibnamefont {Bortolozzo}},\ and\
  \bibinfo {author} {\bibfnamefont {S.}~\bibnamefont {Fauve}},\ }\bibfield
  {journal} {\bibinfo  {journal} {European Physics Letters}\ }\textbf {\bibinfo
  {volume} {86}},\ \href {https://doi.org/10.1209/0295-5075/86/14002}
  {10.1209/0295-5075/86/14002} (\bibinfo {year} {2009})\BibitemShut {NoStop}%
\bibitem [{\citenamefont {Brazhnikov}\ \emph {et~al.}(2002)\citenamefont
  {Brazhnikov}, \citenamefont {Kolmakov}, \citenamefont {Levchenko},\ and\
  \citenamefont {Mezhov-Deglin}}]{brazhnikov_observation_2002}%
  \BibitemOpen
  \bibfield  {author} {\bibinfo {author} {\bibfnamefont {M.~Y.}\ \bibnamefont
  {Brazhnikov}}, \bibinfo {author} {\bibfnamefont {G.~V.}\ \bibnamefont
  {Kolmakov}}, \bibinfo {author} {\bibfnamefont {A.~A.}\ \bibnamefont
  {Levchenko}},\ and\ \bibinfo {author} {\bibfnamefont {L.~P.}\ \bibnamefont
  {Mezhov-Deglin}},\ }\href {https://doi.org/10.1209/epl/i2002-00425-9}
  {\bibfield  {journal} {\bibinfo  {journal} {Europhysics Letters (EPL)}\
  }\textbf {\bibinfo {volume} {58}},\ \bibinfo {pages} {510} (\bibinfo {year}
  {2002})}\BibitemShut {NoStop}%
\bibitem [{\citenamefont {Robinson}(1996)}]{robinson_scalings_1996}%
  \BibitemOpen
  \bibfield  {author} {\bibinfo {author} {\bibfnamefont {P.~A.}\ \bibnamefont
  {Robinson}},\ }\href {https://doi.org/10.1063/1.871845} {\bibfield  {journal}
  {\bibinfo  {journal} {Physics of Plasmas}\ }\textbf {\bibinfo {volume} {3}},\
  \bibinfo {pages} {192} (\bibinfo {year} {1996})}\BibitemShut {NoStop}%
\bibitem [{\citenamefont {Yoon}(2000)}]{yoon_generalized_2000}%
  \BibitemOpen
  \bibfield  {author} {\bibinfo {author} {\bibfnamefont {P.~H.}\ \bibnamefont
  {Yoon}},\ }\href@noop {} {\bibfield  {journal} {\bibinfo  {journal} {Phys.
  Plasmas}\ }\textbf {\bibinfo {volume} {7}},\ \bibinfo {pages} {15} (\bibinfo
  {year} {2000})}\BibitemShut {NoStop}%
\bibitem [{\citenamefont {Boudaoud}\ \emph {et~al.}(2008)\citenamefont
  {Boudaoud}, \citenamefont {Cadot}, \citenamefont {Odille},\ and\
  \citenamefont {Touz{\'e}}}]{boudaoud_observation_2008}%
  \BibitemOpen
  \bibfield  {author} {\bibinfo {author} {\bibfnamefont {A.}~\bibnamefont
  {Boudaoud}}, \bibinfo {author} {\bibfnamefont {O.}~\bibnamefont {Cadot}},
  \bibinfo {author} {\bibfnamefont {B.}~\bibnamefont {Odille}},\ and\ \bibinfo
  {author} {\bibfnamefont {C.}~\bibnamefont {Touz{\'e}}},\ }\href
  {https://doi.org/10.1103/PhysRevLett.100.234504} {\bibfield  {journal}
  {\bibinfo  {journal} {Physical Review Letters}\ }\textbf {\bibinfo {volume}
  {100}},\ \bibinfo {pages} {234504} (\bibinfo {year} {2008})}\BibitemShut
  {NoStop}%
\bibitem [{\citenamefont {Zakharov}\ and\ \citenamefont
  {Filonenko}(1971)}]{zakharov_weak_1971}%
  \BibitemOpen
  \bibfield  {author} {\bibinfo {author} {\bibfnamefont {V.~E.}\ \bibnamefont
  {Zakharov}}\ and\ \bibinfo {author} {\bibfnamefont {N.~N.}\ \bibnamefont
  {Filonenko}},\ }\href {https://doi.org/10.1007/BF00915178} {\bibfield
  {journal} {\bibinfo  {journal} {Journal of Applied Mechanics and Technical
  Physics}\ }\textbf {\bibinfo {volume} {8}},\ \bibinfo {pages} {37} (\bibinfo
  {year} {1971})}\BibitemShut {NoStop}%
\bibitem [{\citenamefont {Zakharov}\ \emph {et~al.}(1992)\citenamefont
  {Zakharov}, \citenamefont {L'vov},\ and\ \citenamefont
  {Falkovich}}]{zakharov_kolmogorov_1992}%
  \BibitemOpen
  \bibfield  {author} {\bibinfo {author} {\bibfnamefont {V.~E.}\ \bibnamefont
  {Zakharov}}, \bibinfo {author} {\bibfnamefont {V.~S.}\ \bibnamefont
  {L'vov}},\ and\ \bibinfo {author} {\bibfnamefont {G.}~\bibnamefont
  {Falkovich}},\ }\href {https://doi.org/10.1007/978-3-642-50052-7} {\emph
  {\bibinfo {title} {Kolmogorov {Spectra} of {Turbulence} {I}}}},\ edited by\
  \bibinfo {editor} {\bibfnamefont {F.}~\bibnamefont {Calogero}}, \bibinfo
  {editor} {\bibfnamefont {B.}~\bibnamefont {Fuchssteiner}}, \bibinfo {editor}
  {\bibfnamefont {G.}~\bibnamefont {Rowlands}}, \bibinfo {editor}
  {\bibfnamefont {H.}~\bibnamefont {Segur}}, \bibinfo {editor} {\bibfnamefont
  {M.}~\bibnamefont {Wadati}},\ and\ \bibinfo {editor} {\bibfnamefont {V.~E.}\
  \bibnamefont {Zakharov}},\ Springer {Series} in {Nonlinear} {Dynamics}\
  (\bibinfo  {publisher} {Springer Berlin Heidelberg},\ \bibinfo {address}
  {Berlin, Heidelberg},\ \bibinfo {year} {1992})\BibitemShut {NoStop}%
\bibitem [{\citenamefont {Connaughton}\ \emph {et~al.}(2001)\citenamefont
  {Connaughton}, \citenamefont {Nazarenko},\ and\ \citenamefont
  {Pushkarev}}]{connaughton_discreteness_2001}%
  \BibitemOpen
  \bibfield  {author} {\bibinfo {author} {\bibfnamefont {C.}~\bibnamefont
  {Connaughton}}, \bibinfo {author} {\bibfnamefont {S.}~\bibnamefont
  {Nazarenko}},\ and\ \bibinfo {author} {\bibfnamefont {A.}~\bibnamefont
  {Pushkarev}},\ }\href {https://doi.org/10.1103/PhysRevE.63.046306} {\bibfield
   {journal} {\bibinfo  {journal} {Physical Review E}\ }\textbf {\bibinfo
  {volume} {63}},\ \bibinfo {pages} {046306} (\bibinfo {year} {2001})},\
  \bibinfo {note} {zSCC: 0000043}\BibitemShut {NoStop}%
\bibitem [{\citenamefont {Falcon}\ \emph {et~al.}(2007)\citenamefont {Falcon},
  \citenamefont {Laroche},\ and\ \citenamefont
  {Fauve}}]{falcon_observation_2007}%
  \BibitemOpen
  \bibfield  {author} {\bibinfo {author} {\bibfnamefont {{\'E}.}~\bibnamefont
  {Falcon}}, \bibinfo {author} {\bibfnamefont {C.}~\bibnamefont {Laroche}},\
  and\ \bibinfo {author} {\bibfnamefont {S.}~\bibnamefont {Fauve}},\ }\href
  {https://doi.org/10.1103/PhysRevLett.98.094503} {\bibfield  {journal}
  {\bibinfo  {journal} {Physical Review Letters}\ }\textbf {\bibinfo {volume}
  {98}},\ \bibinfo {pages} {094503} (\bibinfo {year} {2007})},\ \bibinfo {note}
  {zSCC: 0000201}\BibitemShut {NoStop}%
\bibitem [{\citenamefont {Falcon}\ and\ \citenamefont
  {Laroche}(2011)}]{falcon_observation_2011}%
  \BibitemOpen
  \bibfield  {author} {\bibinfo {author} {\bibfnamefont {E.}~\bibnamefont
  {Falcon}}\ and\ \bibinfo {author} {\bibfnamefont {C.}~\bibnamefont
  {Laroche}},\ }\href {https://doi.org/10.1209/0295-5075/95/34003} {\bibfield
  {journal} {\bibinfo  {journal} {EPL (Europhysics Letters)}\ }\textbf
  {\bibinfo {volume} {95}},\ \bibinfo {pages} {34003} (\bibinfo {year}
  {2011})},\ \bibinfo {note} {arXiv: 1106.3246}\BibitemShut {NoStop}%
\bibitem [{\citenamefont {Deike}\ \emph {et~al.}(2014)\citenamefont {Deike},
  \citenamefont {Berhanu},\ and\ \citenamefont {Falcon}}]{deike_energy_2014}%
  \BibitemOpen
  \bibfield  {author} {\bibinfo {author} {\bibfnamefont {L.}~\bibnamefont
  {Deike}}, \bibinfo {author} {\bibfnamefont {M.}~\bibnamefont {Berhanu}},\
  and\ \bibinfo {author} {\bibfnamefont {E.}~\bibnamefont {Falcon}},\ }\href
  {https://doi.org/10.1103/PhysRevE.89.023003} {\bibfield  {journal} {\bibinfo
  {journal} {Physical Review E}\ }\textbf {\bibinfo {volume} {89}},\ \bibinfo
  {pages} {023003} (\bibinfo {year} {2014})}\BibitemShut {NoStop}%
\bibitem [{\citenamefont {Orosco}\ and\ \citenamefont
  {Friend}(2023)}]{Orosco:2023aa}%
  \BibitemOpen
  \bibfield  {author} {\bibinfo {author} {\bibfnamefont {J.}~\bibnamefont
  {Orosco}}\ and\ \bibinfo {author} {\bibfnamefont {J.}~\bibnamefont
  {Friend}},\ }\href
  {https://authors.elsevier.com/sd/article/S0960-0779(23)00516-7} {\bibfield
  {journal} {\bibinfo  {journal} {Chaos, Solitons \& Fractals}\ }\textbf
  {\bibinfo {volume} {172}},\ \bibinfo {pages} {1} (\bibinfo {year}
  {2023})}\BibitemShut {NoStop}%
\bibitem [{\citenamefont {Kurosawa}\ \emph {et~al.}(1995)\citenamefont
  {Kurosawa}, \citenamefont {Watanabe}, \citenamefont {Futami},\ and\
  \citenamefont {Higuchi}}]{kurosawa_surface_1995}%
  \BibitemOpen
  \bibfield  {author} {\bibinfo {author} {\bibfnamefont {M.}~\bibnamefont
  {Kurosawa}}, \bibinfo {author} {\bibfnamefont {T.}~\bibnamefont {Watanabe}},
  \bibinfo {author} {\bibfnamefont {A.}~\bibnamefont {Futami}},\ and\ \bibinfo
  {author} {\bibfnamefont {T.}~\bibnamefont {Higuchi}},\ }\href@noop {}
  {\bibfield  {journal} {\bibinfo  {journal} {Sensors and Actuators A:
  Physical}\ }\textbf {\bibinfo {volume} {50}},\ \bibinfo {pages} {69}
  (\bibinfo {year} {1995})}\BibitemShut {NoStop}%
\bibitem [{\citenamefont {Collignon}\ \emph {et~al.}(2018)\citenamefont
  {Collignon}, \citenamefont {Manor},\ and\ \citenamefont
  {Friend}}]{collignon_improving_2018}%
  \BibitemOpen
  \bibfield  {author} {\bibinfo {author} {\bibfnamefont {S.}~\bibnamefont
  {Collignon}}, \bibinfo {author} {\bibfnamefont {O.}~\bibnamefont {Manor}},\
  and\ \bibinfo {author} {\bibfnamefont {J.}~\bibnamefont {Friend}},\ }\href
  {https://doi.org/10.1002/adfm.201704359} {\bibfield  {journal} {\bibinfo
  {journal} {Advanced Functional Materials}\ }\textbf {\bibinfo {volume}
  {28}},\ \bibinfo {pages} {1704359} (\bibinfo {year} {2018})}\BibitemShut
  {NoStop}%
\bibitem [{\citenamefont {Lang}(1962)}]{lang_ultrasonic_1962}%
  \BibitemOpen
  \bibfield  {author} {\bibinfo {author} {\bibfnamefont {R.~J.}\ \bibnamefont
  {Lang}},\ }\href {https://doi.org/10.1121/1.1909020} {\bibfield  {journal}
  {\bibinfo  {journal} {The Journal of the Acoustical Society of America}\
  }\textbf {\bibinfo {volume} {34}},\ \bibinfo {pages} {4} (\bibinfo {year}
  {1962})}\BibitemShut {NoStop}%
\bibitem [{\citenamefont {Kurosawa}\ \emph {et~al.}(1997)\citenamefont
  {Kurosawa}, \citenamefont {Futami},\ and\ \citenamefont
  {Higuchi}}]{kurosawa_characteristics_1997}%
  \BibitemOpen
  \bibfield  {author} {\bibinfo {author} {\bibfnamefont {M.}~\bibnamefont
  {Kurosawa}}, \bibinfo {author} {\bibfnamefont {A.}~\bibnamefont {Futami}},\
  and\ \bibinfo {author} {\bibfnamefont {T.}~\bibnamefont {Higuchi}},\ }in\
  \href {https://doi.org/10.1109/SENSOR.1997.635221} {\emph {\bibinfo
  {booktitle} {Proceedings of {International} {Solid} {State} {Sensors} and
  {Actuators} {Conference} ({Transducers} '97)}}},\ Vol.~\bibinfo {volume} {2}\
  (\bibinfo  {publisher} {IEEE},\ \bibinfo {address} {Chicago, IL, USA},\
  \bibinfo {year} {1997})\ pp.\ \bibinfo {pages} {801--804}\BibitemShut
  {NoStop}%
\bibitem [{\citenamefont {Ramisetty}\ \emph {et~al.}(2013)\citenamefont
  {Ramisetty}, \citenamefont {Pandit},\ and\ \citenamefont
  {Gogate}}]{ramisetty_investigations_2013}%
  \BibitemOpen
  \bibfield  {author} {\bibinfo {author} {\bibfnamefont {K.~A.}\ \bibnamefont
  {Ramisetty}}, \bibinfo {author} {\bibfnamefont {A.~B.}\ \bibnamefont
  {Pandit}},\ and\ \bibinfo {author} {\bibfnamefont {P.~R.}\ \bibnamefont
  {Gogate}},\ }\href {https://doi.org/10.1016/j.ultsonch.2012.05.001}
  {\bibfield  {journal} {\bibinfo  {journal} {Ultrasonics Sonochemistry}\
  }\textbf {\bibinfo {volume} {20}},\ \bibinfo {pages} {254} (\bibinfo {year}
  {2013})}\BibitemShut {NoStop}%
\bibitem [{\citenamefont {Qi}\ \emph {et~al.}(2008)\citenamefont {Qi},
  \citenamefont {Yeo},\ and\ \citenamefont {Friend}}]{qi_interfacial_2008}%
  \BibitemOpen
  \bibfield  {author} {\bibinfo {author} {\bibfnamefont {A.}~\bibnamefont
  {Qi}}, \bibinfo {author} {\bibfnamefont {L.~Y.}\ \bibnamefont {Yeo}},\ and\
  \bibinfo {author} {\bibfnamefont {J.~R.}\ \bibnamefont {Friend}},\ }\href
  {https://doi.org/10.1063/1.2953537} {\bibfield  {journal} {\bibinfo
  {journal} {Physics of Fluids}\ }\textbf {\bibinfo {volume} {20}},\ \bibinfo
  {pages} {074103} (\bibinfo {year} {2008})},\ \bibinfo {note} {zSCC:
  0000242}\BibitemShut {NoStop}%
\bibitem [{\citenamefont {Rajan}\ and\ \citenamefont
  {Pandit}(2001)}]{rajan_correlations_2001}%
  \BibitemOpen
  \bibfield  {author} {\bibinfo {author} {\bibfnamefont {R.}~\bibnamefont
  {Rajan}}\ and\ \bibinfo {author} {\bibfnamefont {A.}~\bibnamefont {Pandit}},\
  }\href {https://doi.org/10.1016/S0041-624X(01)00054-3} {\bibfield  {journal}
  {\bibinfo  {journal} {Ultrasonics}\ }\textbf {\bibinfo {volume} {39}},\
  \bibinfo {pages} {235} (\bibinfo {year} {2001})},\ \bibinfo {note} {zSCC:
  0000290}\BibitemShut {NoStop}%
\bibitem [{\citenamefont {Blamey}\ \emph {et~al.}(2013)\citenamefont {Blamey},
  \citenamefont {Yeo},\ and\ \citenamefont {Friend}}]{blamey_microscale_2013}%
  \BibitemOpen
  \bibfield  {author} {\bibinfo {author} {\bibfnamefont {J.}~\bibnamefont
  {Blamey}}, \bibinfo {author} {\bibfnamefont {L.~Y.}\ \bibnamefont {Yeo}},\
  and\ \bibinfo {author} {\bibfnamefont {J.~R.}\ \bibnamefont {Friend}},\
  }\href {https://doi.org/10.1021/la304608a} {\bibfield  {journal} {\bibinfo
  {journal} {Langmuir}\ }\textbf {\bibinfo {volume} {29}},\ \bibinfo {pages}
  {3835} (\bibinfo {year} {2013})}\BibitemShut {NoStop}%
\bibitem [{\citenamefont {Collins}\ \emph {et~al.}(2012)\citenamefont
  {Collins}, \citenamefont {Manor}, \citenamefont {Winkler}, \citenamefont
  {Schmidt}, \citenamefont {Friend},\ and\ \citenamefont
  {Yeo}}]{collins_atomization_2012}%
  \BibitemOpen
  \bibfield  {author} {\bibinfo {author} {\bibfnamefont {D.~J.}\ \bibnamefont
  {Collins}}, \bibinfo {author} {\bibfnamefont {O.}~\bibnamefont {Manor}},
  \bibinfo {author} {\bibfnamefont {A.}~\bibnamefont {Winkler}}, \bibinfo
  {author} {\bibfnamefont {H.}~\bibnamefont {Schmidt}}, \bibinfo {author}
  {\bibfnamefont {J.~R.}\ \bibnamefont {Friend}},\ and\ \bibinfo {author}
  {\bibfnamefont {L.~Y.}\ \bibnamefont {Yeo}},\ }\href
  {https://doi.org/10.1103/PhysRevE.86.056312} {\bibfield  {journal} {\bibinfo
  {journal} {Physical Review E}\ }\textbf {\bibinfo {volume} {86}},\ \bibinfo
  {pages} {056312} (\bibinfo {year} {2012})}\BibitemShut {NoStop}%
\bibitem [{\citenamefont {Barreras}\ \emph {et~al.}(2002)\citenamefont
  {Barreras}, \citenamefont {Amaveda},\ and\ \citenamefont
  {Lozano}}]{barreras_transient_2002}%
  \BibitemOpen
  \bibfield  {author} {\bibinfo {author} {\bibfnamefont {F.}~\bibnamefont
  {Barreras}}, \bibinfo {author} {\bibfnamefont {H.}~\bibnamefont {Amaveda}},\
  and\ \bibinfo {author} {\bibfnamefont {A.}~\bibnamefont {Lozano}},\ }\href
  {https://doi.org/10.1007/s00348-002-0456-1} {\bibfield  {journal} {\bibinfo
  {journal} {Experiments in Fluids}\ }\textbf {\bibinfo {volume} {33}},\
  \bibinfo {pages} {405} (\bibinfo {year} {2002})},\ \bibinfo {note} {zSCC:
  0000175}\BibitemShut {NoStop}%
\bibitem [{\citenamefont {Kooij}\ \emph {et~al.}(2019)\citenamefont {Kooij},
  \citenamefont {Astefanei}, \citenamefont {Corthals},\ and\ \citenamefont
  {Bonn}}]{kooij_size_2019}%
  \BibitemOpen
  \bibfield  {author} {\bibinfo {author} {\bibfnamefont {S.}~\bibnamefont
  {Kooij}}, \bibinfo {author} {\bibfnamefont {A.}~\bibnamefont {Astefanei}},
  \bibinfo {author} {\bibfnamefont {G.~L.}\ \bibnamefont {Corthals}},\ and\
  \bibinfo {author} {\bibfnamefont {D.}~\bibnamefont {Bonn}},\ }\href
  {https://doi.org/10.1038/s41598-019-42599-8} {\bibfield  {journal} {\bibinfo
  {journal} {Scientific Reports}\ }\textbf {\bibinfo {volume} {9}},\ \bibinfo
  {pages} {6128} (\bibinfo {year} {2019})}\BibitemShut {NoStop}%
\bibitem [{\citenamefont {Winkler}\ \emph {et~al.}(2015)\citenamefont
  {Winkler}, \citenamefont {Harazim}, \citenamefont {Menzel},\ and\
  \citenamefont {Schmidt}}]{winkler_saw-based_2015}%
  \BibitemOpen
  \bibfield  {author} {\bibinfo {author} {\bibfnamefont {A.}~\bibnamefont
  {Winkler}}, \bibinfo {author} {\bibfnamefont {S.~M.}\ \bibnamefont
  {Harazim}}, \bibinfo {author} {\bibfnamefont {S.~B.}\ \bibnamefont
  {Menzel}},\ and\ \bibinfo {author} {\bibfnamefont {H.}~\bibnamefont
  {Schmidt}},\ }\href {https://doi.org/10.1039/C5LC00756A} {\bibfield
  {journal} {\bibinfo  {journal} {Lab on a Chip}\ }\textbf {\bibinfo {volume}
  {15}},\ \bibinfo {pages} {3793} (\bibinfo {year} {2015})}\BibitemShut
  {NoStop}%
\bibitem [{\citenamefont {Emery}\ \emph {et~al.}(2021)\citenamefont {Emery},
  \citenamefont {Colomb},\ and\ \citenamefont {Cuche}}]{emery2021metrology}%
  \BibitemOpen
  \bibfield  {author} {\bibinfo {author} {\bibfnamefont {Y.}~\bibnamefont
  {Emery}}, \bibinfo {author} {\bibfnamefont {T.}~\bibnamefont {Colomb}},\ and\
  \bibinfo {author} {\bibfnamefont {E.}~\bibnamefont {Cuche}},\ }\href@noop {}
  {\bibfield  {journal} {\bibinfo  {journal} {Journal of Physics: Photonics}\
  }\textbf {\bibinfo {volume} {3}},\ \bibinfo {pages} {034016} (\bibinfo {year}
  {2021})}\BibitemShut {NoStop}%
\bibitem [{\citenamefont {Berhanu}\ and\ \citenamefont
  {Falcon}(2013)}]{berhanu_space-time-resolved_2013}%
  \BibitemOpen
  \bibfield  {author} {\bibinfo {author} {\bibfnamefont {M.}~\bibnamefont
  {Berhanu}}\ and\ \bibinfo {author} {\bibfnamefont {E.}~\bibnamefont
  {Falcon}},\ }\href {https://doi.org/10.1103/PhysRevE.87.033003} {\bibfield
  {journal} {\bibinfo  {journal} {Physical Review E}\ }\textbf {\bibinfo
  {volume} {87}},\ \bibinfo {pages} {033003} (\bibinfo {year}
  {2013})}\BibitemShut {NoStop}%
\bibitem [{\citenamefont {Taira}\ \emph {et~al.}(2017)\citenamefont {Taira},
  \citenamefont {Brunton}, \citenamefont {Dawson}, \citenamefont {Rowley},
  \citenamefont {Colonius}, \citenamefont {McKeon}, \citenamefont {Schmidt},
  \citenamefont {Gordeyev}, \citenamefont {Theofilis},\ and\ \citenamefont
  {Ukeiley}}]{taira_modal_2017}%
  \BibitemOpen
  \bibfield  {author} {\bibinfo {author} {\bibfnamefont {K.}~\bibnamefont
  {Taira}}, \bibinfo {author} {\bibfnamefont {S.~L.}\ \bibnamefont {Brunton}},
  \bibinfo {author} {\bibfnamefont {S.~T.~M.}\ \bibnamefont {Dawson}}, \bibinfo
  {author} {\bibfnamefont {C.~W.}\ \bibnamefont {Rowley}}, \bibinfo {author}
  {\bibfnamefont {T.}~\bibnamefont {Colonius}}, \bibinfo {author}
  {\bibfnamefont {B.~J.}\ \bibnamefont {McKeon}}, \bibinfo {author}
  {\bibfnamefont {O.~T.}\ \bibnamefont {Schmidt}}, \bibinfo {author}
  {\bibfnamefont {S.}~\bibnamefont {Gordeyev}}, \bibinfo {author}
  {\bibfnamefont {V.}~\bibnamefont {Theofilis}},\ and\ \bibinfo {author}
  {\bibfnamefont {L.~S.}\ \bibnamefont {Ukeiley}},\ }\href
  {https://doi.org/10.2514/1.J056060} {\bibfield  {journal} {\bibinfo
  {journal} {AIAA Journal}\ }\textbf {\bibinfo {volume} {55}},\ \bibinfo
  {pages} {4013} (\bibinfo {year} {2017})}\BibitemShut {NoStop}%
\bibitem [{\citenamefont {Berkooz}\ \emph {et~al.}(1993)\citenamefont
  {Berkooz}, \citenamefont {Holmes},\ and\ \citenamefont
  {Lumley}}]{berkooz_proper_1993}%
  \BibitemOpen
  \bibfield  {author} {\bibinfo {author} {\bibfnamefont {G.}~\bibnamefont
  {Berkooz}}, \bibinfo {author} {\bibfnamefont {P.}~\bibnamefont {Holmes}},\
  and\ \bibinfo {author} {\bibfnamefont {J.~L.}\ \bibnamefont {Lumley}},\
  }\href@noop {} {\bibfield  {journal} {\bibinfo  {journal} {Annual Review of
  Fluid Mechanics}\ }\textbf {\bibinfo {volume} {25}},\ \bibinfo {pages} {539}
  (\bibinfo {year} {1993})}\BibitemShut {NoStop}%
\bibitem [{\citenamefont {Rovira}\ \emph {et~al.}(2021)\citenamefont {Rovira},
  \citenamefont {Engvall},\ and\ \citenamefont {Duwig}}]{rovira_proper_2021}%
  \BibitemOpen
  \bibfield  {author} {\bibinfo {author} {\bibfnamefont {M.}~\bibnamefont
  {Rovira}}, \bibinfo {author} {\bibfnamefont {K.}~\bibnamefont {Engvall}},\
  and\ \bibinfo {author} {\bibfnamefont {C.}~\bibnamefont {Duwig}},\ }\href
  {https://doi.org/10.1103/PhysRevFluids.6.014701} {\bibfield  {journal}
  {\bibinfo  {journal} {Physical Review Fluids}\ }\textbf {\bibinfo {volume}
  {6}},\ \bibinfo {pages} {014701} (\bibinfo {year} {2021})}\BibitemShut
  {NoStop}%
\bibitem [{\citenamefont {Schmidt}\ and\ \citenamefont
  {Colonius}(2020)}]{schmidt_guide_2020}%
  \BibitemOpen
  \bibfield  {author} {\bibinfo {author} {\bibfnamefont {O.~T.}\ \bibnamefont
  {Schmidt}}\ and\ \bibinfo {author} {\bibfnamefont {T.}~\bibnamefont
  {Colonius}},\ }\href {https://doi.org/10.2514/1.J058809} {\bibfield
  {journal} {\bibinfo  {journal} {AIAA Journal}\ }\textbf {\bibinfo {volume}
  {58}},\ \bibinfo {pages} {1023} (\bibinfo {year} {2020})}\BibitemShut
  {NoStop}%
\bibitem [{\citenamefont {Mei}\ \emph {et~al.}(2020)\citenamefont {Mei},
  \citenamefont {Zhang},\ and\ \citenamefont {Friend}}]{Mei:aa}%
  \BibitemOpen
  \bibfield  {author} {\bibinfo {author} {\bibfnamefont {J.}~\bibnamefont
  {Mei}}, \bibinfo {author} {\bibfnamefont {N.}~\bibnamefont {Zhang}},\ and\
  \bibinfo {author} {\bibfnamefont {J.}~\bibnamefont {Friend}},\ }\href
  {https://doi.org/10.3791/61013} {\bibfield  {journal} {\bibinfo  {journal}
  {Journal of Visualized Experiments}\ }\textbf {\bibinfo {volume} {160}},\
  \bibinfo {pages} {e61013} (\bibinfo {year} {2020})}\BibitemShut {NoStop}%
\bibitem [{\citenamefont {Cuche}\ \emph {et~al.}(1999)\citenamefont {Cuche},
  \citenamefont {Marquet},\ and\ \citenamefont
  {Depeursinge}}]{cuche_simultaneous_1999}%
  \BibitemOpen
  \bibfield  {author} {\bibinfo {author} {\bibfnamefont {E.}~\bibnamefont
  {Cuche}}, \bibinfo {author} {\bibfnamefont {P.}~\bibnamefont {Marquet}},\
  and\ \bibinfo {author} {\bibfnamefont {C.}~\bibnamefont {Depeursinge}},\
  }\href {https://doi.org/10.1364/AO.38.006994} {\bibfield  {journal} {\bibinfo
   {journal} {Applied Optics}\ }\textbf {\bibinfo {volume} {38}},\ \bibinfo
  {pages} {6994} (\bibinfo {year} {1999})}\BibitemShut {NoStop}%
\bibitem [{\citenamefont {Shannon}(1949)}]{Shannon:1949aa}%
  \BibitemOpen
  \bibfield  {author} {\bibinfo {author} {\bibfnamefont {C.}~\bibnamefont
  {Shannon}},\ }\href {https://doi.org/10.1109/JRPROC.1949.232969} {\bibfield
  {journal} {\bibinfo  {journal} {Proceedings of the IRE}\ }\textbf {\bibinfo
  {volume} {37}},\ \bibinfo {pages} {10} (\bibinfo {year} {1949})}\BibitemShut
  {NoStop}%
\bibitem [{\citenamefont {Welch}(1967)}]{welch_use_1967}%
  \BibitemOpen
  \bibfield  {author} {\bibinfo {author} {\bibfnamefont {P.}~\bibnamefont
  {Welch}},\ }\href {https://doi.org/10.1109/TAU.1967.1161901} {\bibfield
  {journal} {\bibinfo  {journal} {IEEE Transactions on Audio and
  Electroacoustics}\ }\textbf {\bibinfo {volume} {15}},\ \bibinfo {pages} {70}
  (\bibinfo {year} {1967})}\BibitemShut {NoStop}%
\bibitem [{\citenamefont {Aubourg}\ and\ \citenamefont
  {Mordant}(2016)}]{aubourg_investigation_2016}%
  \BibitemOpen
  \bibfield  {author} {\bibinfo {author} {\bibfnamefont {Q.}~\bibnamefont
  {Aubourg}}\ and\ \bibinfo {author} {\bibfnamefont {N.}~\bibnamefont
  {Mordant}},\ }\href {https://doi.org/10.1103/PhysRevFluids.1.023701}
  {\bibfield  {journal} {\bibinfo  {journal} {Physical Review Fluids}\ }\textbf
  {\bibinfo {volume} {1}},\ \bibinfo {pages} {023701} (\bibinfo {year}
  {2016})}\BibitemShut {NoStop}%
\bibitem [{\citenamefont {Wright}\ \emph {et~al.}(1996)\citenamefont {Wright},
  \citenamefont {Budakian},\ and\ \citenamefont
  {Putterman}}]{wright_diffusing_1996}%
  \BibitemOpen
  \bibfield  {author} {\bibinfo {author} {\bibfnamefont {W.~B.}\ \bibnamefont
  {Wright}}, \bibinfo {author} {\bibfnamefont {R.}~\bibnamefont {Budakian}},\
  and\ \bibinfo {author} {\bibfnamefont {S.~J.}\ \bibnamefont {Putterman}},\
  }\href {https://doi.org/10.1103/PhysRevLett.76.4528} {\bibfield  {journal}
  {\bibinfo  {journal} {Physical Review Letters}\ }\textbf {\bibinfo {volume}
  {76}},\ \bibinfo {pages} {4528} (\bibinfo {year} {1996})},\ \bibinfo {note}
  {zSCC: 0000124}\BibitemShut {NoStop}%
\bibitem [{\citenamefont {Zhang}\ \emph {et~al.}(2021)\citenamefont {Zhang},
  \citenamefont {Orosco},\ and\ \citenamefont {Friend}}]{zhang_onset_2021}%
  \BibitemOpen
  \bibfield  {author} {\bibinfo {author} {\bibfnamefont {S.}~\bibnamefont
  {Zhang}}, \bibinfo {author} {\bibfnamefont {J.}~\bibnamefont {Orosco}},\ and\
  \bibinfo {author} {\bibfnamefont {J.}~\bibnamefont {Friend}},\ }\href
  {http://arxiv.org/abs/2108.09579} {\bibinfo {title} {Onset of low-frequency
  capillary waves driven by high-frequency ultrasound}} (\bibinfo {year}
  {2021}),\ \bibinfo {note} {arXiv:2108.09579 [physics]}\BibitemShut {NoStop}%
\bibitem [{\citenamefont {Manor}\ \emph {et~al.}(2011)\citenamefont {Manor},
  \citenamefont {Dentry}, \citenamefont {Friend},\ and\ \citenamefont
  {Yeo}}]{Manor:2011fk}%
  \BibitemOpen
  \bibfield  {author} {\bibinfo {author} {\bibfnamefont {O.}~\bibnamefont
  {Manor}}, \bibinfo {author} {\bibfnamefont {M.}~\bibnamefont {Dentry}},
  \bibinfo {author} {\bibfnamefont {J.~R.}\ \bibnamefont {Friend}},\ and\
  \bibinfo {author} {\bibfnamefont {L.~Y.}\ \bibnamefont {Yeo}},\ }\href@noop
  {} {\bibfield  {journal} {\bibinfo  {journal} {Soft Matter}\ }\textbf
  {\bibinfo {volume} {7}},\ \bibinfo {pages} {7976} (\bibinfo {year}
  {2011})}\BibitemShut {NoStop}%
\bibitem [{\citenamefont {Pushkarev}\ and\ \citenamefont
  {Zakharov}(1996)}]{pushkarev_turbulence_1996}%
  \BibitemOpen
  \bibfield  {author} {\bibinfo {author} {\bibfnamefont {A.~N.}\ \bibnamefont
  {Pushkarev}}\ and\ \bibinfo {author} {\bibfnamefont {V.~E.}\ \bibnamefont
  {Zakharov}},\ }\href {https://doi.org/10.1103/PhysRevLett.76.3320} {\bibfield
   {journal} {\bibinfo  {journal} {Physical Review Letters}\ }\textbf {\bibinfo
  {volume} {76}},\ \bibinfo {pages} {3320} (\bibinfo {year}
  {1996})}\BibitemShut {NoStop}%
\end{thebibliography}%
% ---------------------------------------------- %
% Appendix
% ---------------------------------------------- %
\appendix

% ---------------------------------------------- %
% Post document
% ---------------------------------------------- %
\section{Welch's method}
We select only the data representing a stationary phenomenon (e.g., we avoid the initial transient) by excluding the first 150,000 time steps or the first $\sim$1.3~s. We divide the remaining time-series into blocks of $2^{15}$ time steps, with a 50\% overlap between blocks. We multiply each block by a Hann window function,
\begin{equation}
    w(m) = (1/2) \big(1-\cos \frac{2\pi m}{M-1}\big),
\end{equation}
where M is the number of time steps in the window and $m=1,2,...,M-1$. We take the periodogram of each block using the discrete Fourier transform, 
\begin{equation}
    S_i(f) = \frac{(\delta t)^2}{T}\big|\sum h_m e^{-i 2 f m \delta t}\big|
\end{equation}
where $\delta t$ is the time step, $8.7 \mu$s, and $T=M*\delta t$, and then take the frequency-wise average over all blocks,
\begin{equation}
    S(f) = \frac{1}{B}\sum S_i(f),
\end{equation}
where B is the number of blocks. Finally we multiply this power spectral density by the frequency spacing and take the square root in order to obtain an amplitude spectrum.

\section{Dynamic Modal Decomposition (DMD)}
To compute the DMD, we choose $N$ frames and construct two matrices, where the frames have been turned into column vectors just as in POD, that are separated by one time step. A matrix, $X_a$, contains frames 1 to $(N-1)$ and a matrix, $X_b$, contains frames 2 to $N$. We perform SVD on $X_a$ and then we construct $X_b$ in terms of an operator that moves $X_a$ forwards in time by one time step. This looks like
\begin{equation}
    X_a = U S V^T
    X_b = A U S V^T
    A~ = U^T X_b V S^(-1),
\end{equation}
where $A~$ is an approximation to the true operator, $A$, but which has the same eigenvalues. We obtain the eigenvalues by
solving $A~ \bf W = \bf W L$ and finally we obtain the eigenvectors of $A$
\begin{equation}
    \bf\Phi = X_b V S \bf W,
\end{equation}
which are the DMD modes. 
\end{document}